\documentclass[preprintnumbers, twocolumn, showpacs, showkeys]{revtex4}
\usepackage{amssymb,amsmath,graphics,graphicx} 
\usepackage[usenames,dvipsnames,svgnames,table]{xcolor}
\usepackage{epstopdf}
\usepackage{enumerate}
\usepackage[utf8]{inputenc}
\usepackage[T1]{fontenc} 
\usepackage[brazil]{babel}

\begin{document}
\title{Demonstração da Lei do Inverso do quadrado com o auxílio de um Tablet/smartphone \\ \small{(Demonstration of the Inverse Square Law with the aid of a Tablet/smartphone)}}
\author{L. P. Vieira$^{\dag}$, V. O. M. Lara$^{\ddag}$ e D. F. Amaral$^{\star}$}
\affiliation{
$^{\dag}$ Instituto de F\'isica - Universidade Federal do Rio de Janeiro, Rio de Janeiro - Rio de Janeiro, Brasil \\
$^{\ddag}$ Instituto de F\'isica - Universidade Federal Fluminense, Niter\'oi - Rio de Janeiro, Brasil \\
$^{\ddag}$ Instituto Federal de Ciência e Tecnologia do Rio de Janeiro, São Gonçalo - Rio de Janeiro, Brasil \\
$^{\star}$ Consórcio de Ensino à distância do Rio de Janeiro (CEDERJ), pólo São Gonçalo -  Rio de Janeiro Brasil  \\
}

\pacs{ 
}

\date{\today}

\begin{abstract}

Neste trabalho mostraremos como obter a lei do inverso do quadrado da distância para a intensidade luminosa emitida por uma fonte pequena de uma maneira muito simples, rápida e com boa precisão. Com o auxílio de dois smartphones, aplicativos grátis e uma régua somos capazes de medir as distâncias entre os dispositivos citados (um como fonte e outro como medidor de intensidade luminosa) e as respectivas intensidades luminosas medidas por um dos aparelhos. A facilidade de reprodução do experimento proposto e a penetração do uso de tablets e smartphones entre os estudantes e professores pode fazer com que a atividade proposta se torne uma boa alternativa para a introdução de fenômenos físicos que exibem a mesma dependência funcional, tais como a Lei da Gravitação Universal de Newton e a Lei de Coulomb. \\

\vspace{0.2 cm}

In this paper we show how to obtain the Inverse-square law of the distance to the light intensity emitted from a small source in a simple, fast and with good precision way. With the aid of two smartphones, free apps and a ruler, one is able to measure the distances between the devices listed (one as a source and the other as a measuring light intensity) and the respective light intensities measured by the probeware in the device. The ease of reproduction of the proposed experiment and the penetration of the use of tablets and smartphones among students and teachers, could make the proposed activity a good alternative to the introduction of physical phenomena which exhibit the same functional dependence, such as the Law of Newton's Universal Gravitation and Coulomb's Law.

\end{abstract}


\maketitle

\section*{Introdução}

Sabemos que quando as dimensões de uma fonte luminosa são pequenas quando comparadas à distância entre essa fonte e um receptor dessa luz, podemos considerar que a fonte luminosa em questão é pontual. Uma maneira de confirmar os limites de validade dessa aproximação é medindo a queda da intensidade de luz que chega a um determinado “ponto”, aumentando-se a distância entre a fonte e o medidor e mantendo a potência luminosa da fonte fixa. Se a intensidade luminosa cair com o inverso do quadrado desta distância, podemos considerar que a fonte é pontual. Pode-se argumentar também que essa fonte emite luz em todas as direções igualmente, ou seja, a única simetria que permite essa configuração é a esférica, com a fonte luminosa localizada na posição central.

\begin{figure}[!htb]
\begin{center}
\includegraphics[scale=0.56]{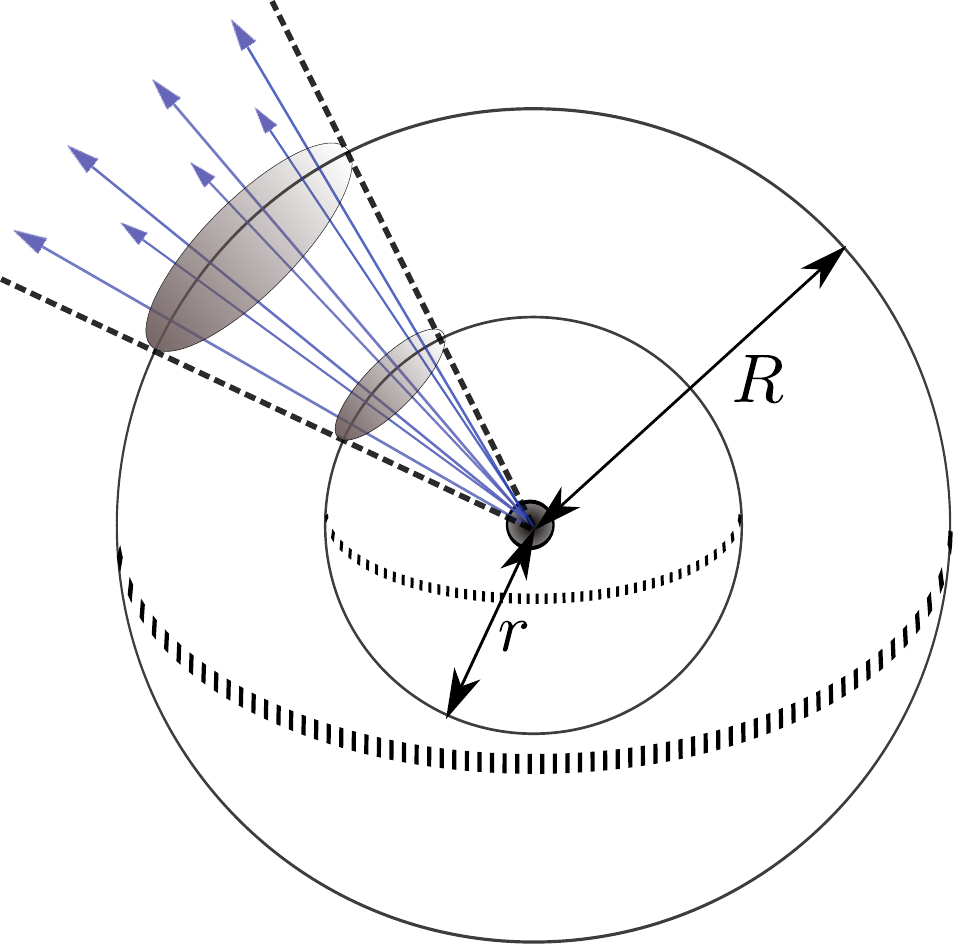}
\end{center}
\caption{Representação esquemática de um cone de luz advindo de uma fonte que emite igualmente em todas as direções. Repare que a densidade superficial de raios luminosos cai com a distância ao quadrado, conforme discutido no texto. Utilizamos sete raios por simplicidade de representação, uma vez que a quantidade de raios luminosos não é finita. Mesmo assim, a densidade de raios diminui com a mesma forma funcional.}
\label{fig:fonte_luz}
\end{figure}

Podemos ainda nos utilizar de argumentos mais rigorosos. Usando o modelo de raios luminosos, dizemos que toda e qualquer região fechada ao redor dessa fonte receberá um certo fluxo de raios luminosos, que independe da distância, de maneira análoga ao que se faz com as linhas de campo elétrico emanadas por uma carga elétrica pontual \cite{griffiths}. A quantidade de energia que chega a uma certa região infinitesimal depende diretamente da densidade de raios luminosos, da área desta região e do produto escalar entre o vetor área (normal à região) e o vetor $\hat{k}$ referente à direção de propagação dos raios. Com a ideia de densidade de raios luminosos, definida como a quantidade de raios que atravessam uma determinada área dividido pelo valor desta área, vemos que a densidade de raios luminosos que se propagam dentro de um dado ângulo sólido formado a partir da fonte diminui à medida que nos afastamos da fonte. Isto ocorre por que a área aumenta e a quantidade de raios luminosos se mantém a mesma [veja a figura (\ref{fig:fonte_luz})].

Como a área da esfera aumenta com $r^{2}$ (onde $r$ é a distância entre a fonte e o medidor), temos, para o caso de uma fonte pontual, que a densidade de raios luminosos - e também a intensidade luminosa - é proporcional a $r^{-2}$, ou seja:

\begin{equation}
I \propto r^{-2} .
\label{eq:1}
\end{equation} 

Esse é um assunto de grande relevância na Física, principalmente no entendimento de alguns modelos, entre os quais ressaltamos a Gravitação Universal de Newton e a Lei de Coulomb. Em ambos os casos os campos de partículas pontuais também exibem esse comportamento ($E \propto r^{-2}$), excetuando-se distribuições mais complexas de carga (dipolos e multipolos de ordem superior).

Para discutir este tema introduzimos o uso de smartphones, aparelhos que fazem parte do cotidiano dos alunos. Há trabalhos anteriores nesta direção. Para os interessados, recomendamos algumas referências \cite{artigo_int1, artigo_int2, artigo_gota, artigo_ondas, diss_leo}.

\section*{Os smartphones como medidores de grandezas físicas}

Os smartphones e tablets, tão difundidos hoje em dia, podem servir como computadores pessoais e instrumento de medida direta de grandezas físicas importantes no Ensino de Física. Sensores e softwares (aplicativos) de fácil acesso e utilização já vêm de fábrica nesses aparelhos, podendo ser usados para medir aceleração, velocidade angular, intensidade sonora, campo magnético, posição (através de GPS) e intensidade luminosa. Aplicativos podem ser baixados para efetuar a leitura, armazenamento e apresentação dos dados mensurados.

Para medir a intensidade de luz usaremos o luxímetro, localizado na câmera de smartphones e tablets. Vários aplicativos de medição de intensidade luminosa podem ser baixados gratuitamente para as três principais plataformas (Android, iOS e WindowsPhone) \cite{android, ios, windowsphone}.
	
\begin{figure}[!htb]
\begin{center}
\includegraphics[scale=0.65]{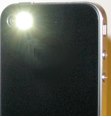}
\includegraphics[scale=0.09]{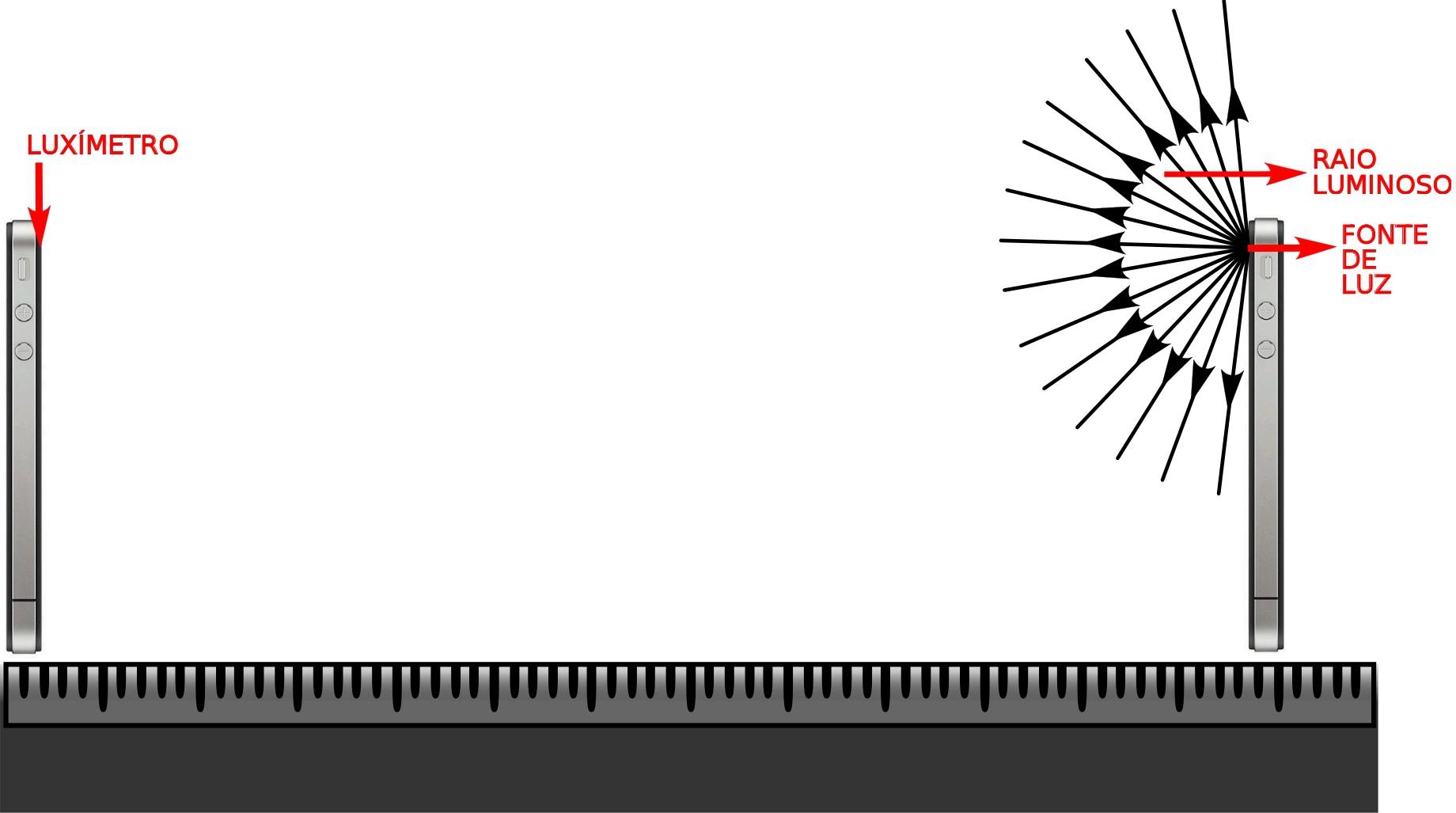}
\end{center}
\caption{À esquerda temos a fotografia de um smartphone que está com o flash ativado. À direita temos uma representação esquemática do experimento descrito no texto.}
\label{fig:fonte_celular}
\end{figure}

\begin{figure}[!htb]
\begin{center}
\vspace{0.6cm}
\includegraphics[scale=0.27]{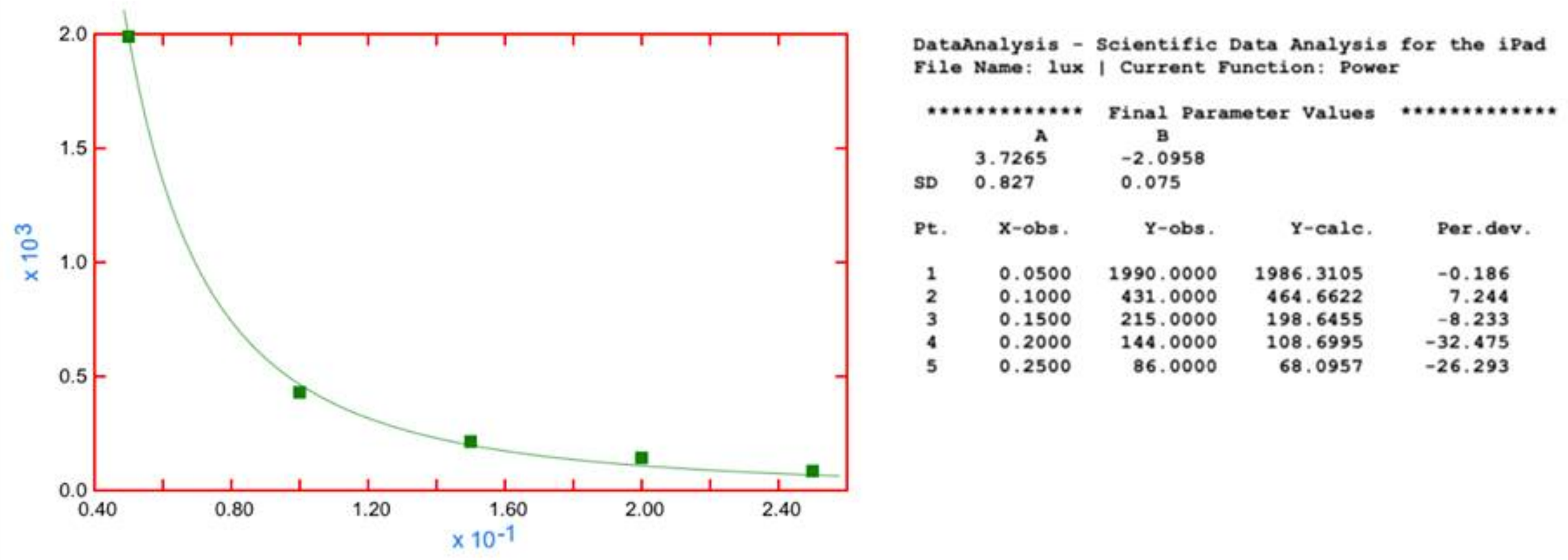}
\end{center}
\caption{Nesta figura temos o gráfico da intensidade luminosa em função da distância entre fonte e sensor, bem como os valores dos parâmetros $a = 3.72 \pm 0.82$ e $b = -2.09 \pm 0.07$ obtidos para o ajuste utilizando-se a função $f(x) = ax^{b}$.}
\label{fig:grafico}
\end{figure}

Uma vez que você tenha instalado devidamente um desses aplicativos em um smartphone ou tablet você será capaz de medir com boa precisão a intensidade luminosa que chega à câmera digital desse aparelho. Agora, utilizando a luz do flash de um segundo aparelho como fonte luminosa \footnote{A maioria desses aparelhos permitem que se utilize o LED de flash como lanterna [veja a figura (\ref{fig:fonte_celular})] e nada o impede de usar uma lâmpada comum}, afixamos um desses aparelhos e alteramos a distância entre eles, anotando os valores da intensidade luminosa e distância correspondentes [veja as  figuras (\ref{fig:fonte_celular}) e (\ref{fig:grafico})].

O gráfico da figura (\ref{fig:grafico}) foi gerado com um aplicativo grátis baixado na \textit{Applestore} \cite{dataanalysis}. Esse é capaz de plotar um conjunto de pontos e ajustar uma curva a estes pontos com bastante rapidez. Podemos observar que o ajuste foi realizado com uma função lei de potência do tipo $y = ax^{b}$. O expoente $b$ calculado, para os valores medidos, é por volta de -2 com ótima precisão até a primeira casa decimal. Isto significa que as distâncias utilizadas no experimento são grandes em comparação com o tamanho da fonte, e portanto, podemos considerar esse tipo de fonte luminosa como pontual. Vale ressaltar que mesmo com luminosidade de fundo os resultados mostrados aqui ainda são reprodutíveis, contanto que você encontre para que distância a luminosidade da fonte tem intensidade parecida com a de fundo, tomando-a como limite superior. 

Caso o interessado deseje reproduzir este experimento utilizando uma lâmpada incandescente, deverá ter o cuidado de começar a coleta de dados para distâncias um pouco maiores. Sugerimos que o comprimento do filamento de tungstênio seja estimado e que se comecem as medições para distâncias maiores que o triplo do tamanho estimado do filamento.

\section*{Conclusões e Perspectivas}

Observamos que essa técnica simples e de fácil acesso pode ajudar os alunos a compreender que a simetria com a qual um sistema físico se dispersa ou propaga tem grande influência na maneira como o fenômeno se comporta. A simetria esférica se apresenta em vários modelos físicos e um experimento como esse é importante nas interações de aprendizagem. 
	
	Utilizar-se de uma lente convergente para colimar esse feixe luminoso e repetir o experimento mostrará  que o comportamento da luminosidade, quando comparada com a distância da fonte, se alterará fortemente. Dessa vez a intensidade não diminui ou talvez, diminuirá muito pouco. Temos aí um ótimo exemplo de como a simetria dos processos físicos alteram seu comportamento. Se logo após a lente esférica, posicionarmos uma cilíndrica, devemos perceber uma queda linear com o aumento da distância. São muitas as possibilidades que um arranjo simples e prático como esse nos proporciona.      

\section*{Agradecimentos}

Os autores são gratos à Agência de Fomento CAPES.

\bibliographystyle{apsrev.bst}
\bibliography{mybib.bib} 

\end{document}